\documentclass[sigconf,natbib=true,authorversion=true,review=false,anonymous=false]{acmart}

\usepackage{xcolor}
\usepackage{tabularx}
\usepackage{acronym}
\acrodef{AI}[AI]{Artificial Intelligence}
\acrodef{AGI}[AGI]{Artificial General Intelligence}
\acrodef{CHM}[CHM]{Collaborative-Human-Machine}
\acrodef{CS}[CS]{Conversational Search}
\acrodef{IR}[IR]{Information Retrieval}
\acrodef{LLM}[LLM]{Large Language Model}
\acrodef{ML}[ML]{Machine Learning}
\acrodef{QA}[QA]{Question Answering}
\acrodef{QPP}[QPP]{Query Performance Prediction}

\usepackage{color}
\definecolor{ForestGreen}{rgb}{0.13, 0.55, 0.13}

\newcommand{\manualJudgment}[0]{human judgment}
\newcommand{\aiAssistance}[0]{AI assistance}
\newcommand{\manualVerification}[0]{human verification}
\newcommand{\fullyAutomated}[0]{fully automated}

\newcommand{\manualJudgmentTitle}[0]{Human Judgment}
\newcommand{\aiAssistanceTitle}[0]{AI Assistance}
\newcommand{\manualVerificationTitle}[0]{Human Verification}
\newcommand{\fullyAutomatedTitle}[0]{Fully Automated}

\newcommand{\mycaption}[1]{\caption{#1}}
\usepackage{balance}
\usepackage{xurl}
\usepackage{titlecaps}
\usepackage{enumitem}
\usepackage{mdframed}
\usepackage{multirow}
\usepackage{rotating}

\definecolor{lightgrey}{RGB}{240,240,240}

\newmdenv[
  linewidth=1pt,
  linecolor=black,
  backgroundcolor=lightgrey,
  roundcorner=5pt,
  leftmargin=1cm,
  rightmargin=1cm,
]{greyQuote}

\graphicspath{{./figures/}}

\begin{document}

\title{Perspectives on Large Language Models for Relevance Judgment}


\author{Guglielmo Faggioli}
\orcid{}
\affiliation{%
  \institution{University of Padova}
  \city{}
  \country{}
}
\email{}

\thanks{We thank Ian Soboroff for his ideas, comments, and other contributions.}

\author{Laura Dietz}
\orcid{https://orcid.org/0000-0003-1624-3907}
\affiliation{%
  \institution{University of New Hampshire}
  \city{}
  \country{}
}
\email{}

\author{Charles L.\ A.\ Clarke}
\orcid{}
\affiliation{%
  \institution{University of Waterloo}
  \city{}
  \country{}
}
\email{}

\author{Gianluca Demartini}
\orcid{}
\affiliation{%
  \institution{University of Queensland}
  \city{}
  \country{}
}
\email{}

\author{Matthias Hagen}
\orcid{}
\affiliation{%
  \institution{Friedrich-Schiller-Universit{\"a}t Jena}
  \city{}
  \country{}
}
\email{}

\author{Claudia Hauff}
\orcid{}
\affiliation{%
  \institution{Spotify}
  \city{}
  \country{}
}
\email{}

\author{Noriko Kando}
\orcid{}
\affiliation{%
  \institution{National Institute of Informatics (NII)}
  \city{}
  \country{}
}
\email{}

\author{Evangelos Kanoulas}
\orcid{}
\affiliation{%
  \institution{University of Amsterdam}
  \city{}
  \country{}
}
\email{}

\author{Martin Potthast}
\orcid{}
\affiliation{%
  \institution{Leipzig University and ScaDS.AI}
  \city{}
  \country{}
}
\email{}


\author{Benno Stein}
\orcid{}
\affiliation{%
  \institution{Bauhaus-Universit{\"a}t Weimar}
  \city{}
  \country{}
}
\email{}

\author{Henning Wachsmuth}
\orcid{}
\affiliation{%
  \institution{Leibniz University Hannover}
  \city{}
  \country{}
}
\email{}

\copyrightyear{2023}
\acmYear{2023}
\acmConference{April 2023}{arXiv}{Internet}
\acmBooktitle{}
\acmDOI{}
\acmISBN{}
\settopmatter{printacmref=false}

\begin{abstract}
When asked, current large language models~(LLMs) like ChatGPT claim that they can assist us with relevance judgments. Many researchers think this would not lead to credible IR research. 
In this perspective paper, 
we discuss possible ways for LLMs to assist human experts along with concerns and issues that arise.
We devise a human--machine collaboration spectrum that allows categorizing different relevance judgment strategies, based on how much the human relies on the machine. For the extreme point of "fully automated assessment", we further include a pilot experiment on whether LLM-based relevance judgments correlate with judgments from trained human assessors. We conclude the paper by providing two opposing perspectives—for and against the use of LLMs for automatic relevance judgments—and a compromise perspective, informed by our analyses of the literature, our preliminary experimental evidence, and our experience as IR researchers. 

We hope to start a constructive discussion within the community to avoid a stale-mate during review, where work is dammed if is uses LLMs for evaluation and dammed if it doesn't.
\end{abstract}



\maketitle

\renewcommand{\shortauthors}{Faggioli, Dietz, Clarke, Demartini, Hagen, Hauff, Kando, Kanoulas, Potthast, Stein, and Wachsmuth}

\section{Introduction} \label{sec:introduction}

That evaluation is very important to the information retrieval~(IR) community is demonstrated by long-standing evaluation campaigns spread throughout the world~\cite{FIRE2022,Harman1993,Kando1999,Braschler2001}. The difficulty of a proper evaluation setup in~IR is also well-known~\cite{Saracevic1995,Jarvelin2009,Fuhr2018,Sakai2021}. IR~evaluation traces its roots back to the Cranfield paradigm~\cite{Cleverdon1960}, which is based on the concept of test collections consisting of (i)~a document corpus, (ii)~a set of information needs or topics, and (iii)~relevance judgments for documents on the topics. Critically, according to the Cranfield paradigm, human assessors are needed for the relevance judgments---a time-intensive and costly procedure.%
\footnote{As a concrete example, for the 50~topics in the TREC-8 Ad Hoc track~\cite{VoorheesHarman1999}, 129~participating systems led to more than 86,000~pooled documents to judge, requiring more than 700~assessor hours at a cost of about~USD~15,000.}

\begin{figure}
\begin{greyQuote}
\colorbox{white}{
\includegraphics{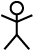} Dear\,LLM, can\,you\,assist\,us\,with\,relevance\,judgments?}

\ \,\includegraphics{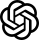} Yes, I can certainly provide assistance with relevance judgments. Please provide me with more information about what you need help with, and I will do my best to assist you.

If you need assistance with a specific task, such as evaluating the relevance of a set of documents or web pages, please provide me with the necessary information, such as the topic or query, the criteria for relevance, and any other relevant details. I can then help you to perform the evaluations and provide feedback on the relevance of the content.
\end{greyQuote}

\caption{Asking ChatGPT for assistance on Feb. 15, 2023.}
\label{fig:ask-chatgpt}
\end{figure}

However, over the past decades, we have become used to witnessing tasks that were traditionally performed by humans being delegated to machines, starting with indexing and retrieval. While the idea of automatically generated judgments~\cite{soboroff2001ranking} has been considered before, it has not found widespread use in the IR~community. Other routes to minimize the cost of collecting relevance judgments in the past include judging text nuggets instead of documents~\cite{pavlu2012ir}, using crowdworkers~\cite{alonso2009can,BlancoHalpinEtAl2011} (though this comes with its own set of problems~\cite{maddalena2016crowdsourcing}), cleverly selecting which documents to judge~\cite{CarteretteAllanEtAl2006,jayasinghe2014improving}, constructing test collections from Wikipedia~\cite{dietz2022wikimarks}, or automating parts of the judgment process via a QA~system~\cite{sander2021exam}.

Figure~\ref{fig:ask-chatgpt} shows the response of ChatGPT%
\footnote{\url{https://chat.openai.com/chat}}
when asked if it can assist with relevance judgments. The response suggests that it is able to carry out relevance judgments, but it is unclear how well such judgments align with those made by human annotators. In this perspectives paper, we explore whether we are on the verge of being able to delegate the process of relevance judgment to machines too, by means of large language models~(LLMs)---either fully or partially, across different domains and tasks or just for a select few. We aim to provide a balanced view on this contentious statement by presenting both consenting and dissenting voices in the scientific debate surrounding the use of~LLMs for this purpose. Although a variety of document modalities exist (audio, video, images, text), we here focus on text-based test collections. We opt for text collections being the most commonly used ones in~IR: the consolidated methodology for assessing the relevance of textual documents, which dates back to the Cranfield paradigm, enables us to carry a ground comparison between~LLMs and human assessors.

While the technology might not be ready yet to provide fully automatic relevance judgments, we argue that LLMs are already able to help humans in this task---to various extents. To model the range of automation, we propose a spectrum illustrating the degrees of collaboration between humans and~LLMs (see Table~\ref{table-perspective-collaboration}). This spectrum spans from manual judgments, the current setup, to fully automated judgments that are carried out solely by~LLMs, a potentially envisioned perspective. The level of human involvement and decision making varies along the spectrum.

\paragraph{Contributions.}
In this perspectives paper, we discuss a spectrum of scenarios of leveraging human--machine collaboration for relevance judgments in IR contexts. Some scenarios have been studied already and are elaborated in the related work section. Others are currently emerging, for which we describe risks as well as open questions that require further research. We also conduct a pilot feasibility experiment where we assess to what extent judgments generated by~LLMs agree with human judgments, including an analysis of LLM-specific caveats. To conclude our paper, we provide two opposing perspectives---for and against the use of~LLMs for automatic relevance judgments---as well as a compromise between them. All of them are informed by our analyses of the literature, our pilot experimental evidence, and our experience as IR~researchers.
\section{Related Work}

The test collection approach to \ac{IR} requires the creation of queries, documents and relevance judgments to be created. The traditional approach is to hire human assessors, to provide relevant judgments. However, the manual effort associated with their creation is staggering, leading to a range of approaches to either assist the assessor or automate tedious tasks. The goal is to both improve the annotation quality, consistency, and efficiency of the assessment. Below we describe existing approaches and relate them to the Human-Machine-Collaboration spectrum.

\subsection{\manualJudgmentTitle{}}



\paragraph{Assessment Systems} 
\citet{NevesSeva2021} provide a rich survey of tools used by human experts in annotating documents. They identify a set of 13 features of such tools that help the human assessor in completing their task, such as text highlighting support for pre-annotations and integration with external data sources (e.g., ontologies and thesauri)~\cite{StenetorpPyysaloEtAl2012,YimamGurevych2013}.

\paragraph{Crowdsourcing} As document collections kept growing in size, the ratio of documents that could practically be judged by human assessors kept getting smaller. This triggered the research community to look for ways to scale-up the collection of human-generated relevance judgments.
Around 2010, research looking at replacing trained human assessors leveraging micro-task crowdsourcing started to appear \cite{alonso2009can}. In the last 10 years, the community has been looking at research questions related to the reliability of crowdsourced relevance judgments \cite{BlancoHalpinEtAl2011} as well as at questions related to cost and quality management \cite{maddalena2016crowdsourcing}. The increase in possible scale and accessibility of work power usually comes with a decrease in reliability, often due to the complicated interaction of crowd workers and task requesters \cite{nouri:2020}.
The current understanding based on research findings is that crowdsourcing relevance judgments is a viable solution to scaling up the collection of labels and an alternative approach to traditional relevance judgments performed by trained human assessors. This is true as long as quality control mechanisms are put in place and the domain is accessible to non-experts~\cite{tamine2017impact}. Quality control mechanisms may include label aggregation methods \cite{sheshadri2013square}, task design strategies \cite{AlonsoBaezayates2011,kazai2013analysis}, and crowd worker selection strategies \cite{gadiraju2019crowd}.
Recent research has looked at how to support crowd workers in judging relevance by presenting them with extra information (e.g., machine-generated metadata) that can increase their judgment efficiency \cite{xu2023role}.

\subsection{\manualVerificationTitle{} and \aiAssistanceTitle{}}

In this scenario, the human partially relinquishes control over which documents will be assessed or how the assessments will be derived by the machine but remains in control of defining relevance.

\paragraph{Passage-ROUGE and BertScore.} As a cost-effective means to judge passages, \citet{keikha2014evaluating} expand automatically manual relevance judgments: any unjudged passage that has a high similarity to a judged passage, will inherit its relevance label. They explore the ROUGE measure as a similarity. Alternatively, approaches such as BertScore \cite{zhang2019bertscore} can serve as an LLM-based similarity.

\paragraph{AutoTar.} Several approaches to semi-automatic support in test collection creation have been proposed. One approach is to use active learning for annotation \cite{cormack1998efficient}, where the pool of documents to manually assess is determined based on the confidence of a machine learning algorithm, the role of the human is to assess given documents. 
This approach is very successful when the
failure to identify a relevant document must be avoided.
Similarly, \citet{jayasinghe2014improving} describe a method for selecting documents to be included in a test collection using a machine learning approach: the proposed methodology finds relevant documents that would otherwise only be found using manual runs, and allows for constructing a low-bias reusable test collection.

\paragraph{Estimating AP} Alternatively, evaluation metrics can be adjusted to correct for biases of incomplete judgments \cite{yilmaz2008simple}. This approach reduces the cost by reducing the number of assessments needed for evaluating search systems. 


 \paragraph{EXAM} Instead of asking humans to assess each document for relevance, for the EXAM Answerability Metric \citet{sander2021exam} ask humans to design a set of exam questions that can be answered with relevant text. An automatic question-answering System is asked to answer these exam questions by using the content of retrieved documents.  The idea is that the more questions can be answered correctly with the document, the better the search system that retrieved the document is. Similar paradigms have been used successfully in other labeling tasks as well  \cite{eyal2019question,deutsch2021towards,huang2020knowledge}

\paragraph{Query Performance Prediction} A related body of work concerns the \ac{QPP}, which is defined as the task of evaluating the performance of an \ac{IR} system, in the absence of human-made relevance judgements~\cite{Hauff2010,CarmelYomtov2010}. 
In this regard, our proposal for automatic assessment of the documents using \ac{LLM} not only would provide benefits to a number of downstream tasks, such as \ac{QPP}, but its effectiveness has already been partially shown and is supported by flourishing literature concerning \acp{LLM} in the \ac{QPP} domain~\cite{ArabzadehKhodabakhshEtAl2021,DattaMacavaneyEtAl2022,ChenHeEtAl2022,ArabzadehSeifikarEtAl2022}.

\subsection{\fullyAutomatedTitle{} Test Collections}
A further strategy to devise queries automatically is the Wikimarks approach \citep{dietz2022wikimarks}. Wikimarks derives queries from the title and heading structure of Wikipedia articles, with passages below taken as relevant. This approach has also been applied to aspect-based summarization \cite{hayashi2021wikiasp}, and text segmentation \cite{ArnoldSchneiderEtAl2019}.

\paragraph{Reconstruct Documents.} Instead of hiring assessors, repositories of semi-structured (human-authored) articles can be used to derive what the human author considered relevant. To this end approaches use anchor text \citep{AsadiMetzlerEtAl2011}, metadata of scientific article sections \citep{BerendsenTsagkiasEtAl2012}, categories in the Open Directory Project \citep{BeitzelJensenEtAl2003},  glosses in Freebase \citep{dalvi2015automatic} or infoboxes \cite{kasneci2009yago,hewlett2016wikireading}.

\paragraph{Evaluation of Automatic Evaluation.} A question is how well automatic assessments would agree with manual assessments. To this end, a study on the correlation of leaderboards on the TREC CAR data found a very high-rank correlation \citep{dietz2020humans}. We repeat a similar study in the context of LLMs in Section \ref{sec:evaluation}.

\section{Spectrum of Human--Machine Collaboration} \label{sec:spectrum}

\providecommand{\bscell}{}
\providecommand{\bscellA}{}
\providecommand{\bscellB}{}

\renewcommand{\bscell}[2]{%
  \parbox[t]{#1\textwidth}{\raggedright #2\rule[-1ex]{0em}{0ex}}}

\renewcommand{\bscellA}[2][0.11]{\bscell{#1}{#2}}
\renewcommand{\bscellB}[2][0.32]{\bscell{#1}{#2}}

\newlength\scaledlength

\newcommand{\humanvsllm}[2][]{{%
  \setlength\scaledlength{0.25pt}%
  \multiply\scaledlength by #2
  \renewcommand{\tabcolsep}{2pt}%
  \renewcommand{\arraystretch}{1}
  \begin{tabular}{clc}
  \includegraphics{human-logo} & \rule[6pt]{25pt}{0.4pt}\kern4pt & \kern-4pt\includegraphics{chatgpt-logo}#1\\[-12pt]
  & \kern-3pt\hspace{\scaledlength}$\triangle$
  \end{tabular}}}

\begin{table}[t]
\centering
\setlength{\tabcolsep}{0pt}

\mycaption{A spectrum of collaborative 
\protect\includegraphics[height=\fontcharht\font`T]{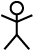} human -- 
\protect\includegraphics[height=\fontcharht\font`T]{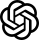} machine task organization to produce relevance judgments. The $\triangle{}$ indicates where on the spectrum each possibility falls.}
\label{table-perspective-collaboration}

\resizebox{\columnwidth}{!}{
\begin{tabular}{l @{\quad} c}
\toprule

\bscellA{\centering\bfseries Collaboration Integration} & \raisebox{-2ex}{\bfseries Task Organization} \\
\midrule

\titlecap{Human Judgment} \\[0.5ex]
\humanvsllm{0} &
\bscellB{Humans do all judgments manually without any kind of support.} \\
\addlinespace

\humanvsllm{10} &
\bscellB{Humans have full control of judging but are supported by text highlighting, document clustering, etc.} \\
\addlinespace

\titlecap{AI Assistance} \\[0.5ex]
\humanvsllm{30} &
\bscellB{Humans judge documents while having access to LLM-generated summaries.} \\
\addlinespace

\humanvsllm{50} &
\bscellB{Balanced competence partitioning. Humans and LLMs focus on (sub-)tasks they are good at.}\\
\addlinespace

\titlecap{Human Verification} \\[0.5ex]
\humanvsllm[\includegraphics{chatgpt-logo}]{67} &
\bscellB{Two LLMs each generate a judgment, and humans select the better one.} \\
\addlinespace

\humanvsllm{77} &
\bscellB{An LLM produces a judgment (and an explanation) that humans can accept or reject.} \\
\addlinespace

\humanvsllm[\raisebox{2pt}{$\,\cdot n$}]{77} &
\bscellB{LLMs are considered crowdworkers with varied specific characteristics, but supervised~/~controlled by humans.} \\
\addlinespace

\titlecap{Fully Automated} \\[0.5ex]
\humanvsllm{100} &
\bscellB{Fully automatic judgments.} \\
\addlinespace

\bottomrule
\end{tabular}}

\end{table}


To identify what contributions LLMs may provide to relevance judgments, we devise a human--machine collaboration spectrum. This spectrum outlines different levels of collaboration between humans and LLMs. At one end, humans make judgments manually, while at the other end, LLMs replace humans completely. In between, LLMs assist humans with various degrees of interdependence. A summary of our proposed four levels of human--machine collaboration is shown in Table~\ref{table-perspective-collaboration}. In the following, we discuss each level in detail.

\paragraph{\titlecap{\manualJudgment{}}} On one extreme, humans do all judgments manually and decide what is relevant without being influenced by an~LLM. In reality, of course, humans are still supported with basic features of the judgment interface. Such features might be based on heuristics, but should not require any form of automatic training/feedback. For instance, humans may define ``scan terms'' to be highlighted in the text, they may limit viewing the pool of documents that have already been judged, or they may order documents so that similar documents are near each other. This end of the spectrum thus represents the status quo, where humans are, in the end, the only reliable judges.
    
\paragraph{\titlecap{\aiAssistance{}}}Advanced assistance can come in many forms, for example, an~LLM may generate a summary of a to-be-judged document so that the human assessor can more efficiently make a judgment based on this compressed representation. Another approach could be to manually define information nuggets that are relevant (e.g., exam questions~\cite{sander2021exam}) and to then train an~LLM to automatically determine how many test nuggets are contained in the retrieved results (e.g., via a QA~system).

This leads us to the first research direction towards improving the human--machine collaboration: \emph{How to employ LLMs, as well as other AI~tools, to aid human assessors in devising reliable judgments while enhancing the efficiency of the process?} What are tasks that can be taken over by~LLMs (e.g., document summarization or keyphrase extraction)? 
    
\paragraph{\titlecap{\manualVerification{}}} For each document to judge, a first-pass judgment of an~LLM is automatically produced as a suggestion along with a generated rationale. We consider this to be a \textit{human-in-the-loop} approach: one or more~LLMs provide their relevance judgment and the human verifies them. In most cases, the human will therefore be assigned menial and undemanding tasks, or will not have to intervene at all. Regardless, the human might still be required in challenging scenarios or situations where the~LLM has low confidence. Another approach could follow the ``preference testing'' paradigm~\cite{windsor1994preference} where two machines each generate a judgment, and a human will select the better one---intervening only in the case of disagreements between the machines and verifying the information. In both cases, humans make the ultimate decision wherever needed. The concern is that any bias in the~LLM might be affecting relevance judgments, as humans will not be able to correct for information they will not see. 

Concerning this layer of the spectrum, the research direction that we wish to raise within the community is: What sub-tasks of the judgment process require human input (e.g., prompt engineering~\cite{ZhouMuresanuEtAl2022,SorensenRobinsonEtAl22}---for now) and \emph{for what tasks should human assessors not be replaced by machines?}

\paragraph{\titlecap{\fullyAutomated{}}} If LLMs were able to assess relevance reliably, they could completely replace humans in the judgment process. We explore the possibility that a fully automatic judgment system might be as good as a human in producing high-quality relevance judgments (for a specific corpus/domain). Automatic judgments might even surpass the human in terms of quality, which raises follow-up issues (cf.\ Section~\ref{sec:beyond-human}).

In this regard, the third research direction that our community should investigate is: \emph{How can  humans be replaced entirely by~LLMs in the judgment process?} Indeed, one can go as far as asking whether generative~LLMs can be used to create new test collections by creating new corpora, queries, abstracts, and conversations.

\medskip
A central aspect to be investigated is where on this four-level human--machine collaboration spectrum we actually obtain the ideal relevance judgments at the best cost. At this point, humans perform tasks that humans are good at (maybe none?!), while machines perform tasks that machines are good at. We refer to this scenario as \textit{competence partitioning}~\cite{hancock2013task,flemisch2016shared}: the task is assigned to either the human or the machine, depending on who is better. Note that in our current version of the spectrum, we still (optimistically) show balanced competence partitioning as part of ``AI assistance''.
\section{Open Issues, Foreseeable Risks, and Opportunities} \label{sec:dangers}

In this section, we look at different issues that come up when \acp{LLM} are used for relevance judgment tasks. We discuss open research questions, risks we foresee, as well as opportunities to move beyond the currently accepted \ac{IR} evaluation paradigms.

\subsection{AI Assistance}

\providecommand{\bscellR}{}
\renewcommand{\bscellR}[1]{%
  \kern-0.5em\makebox[2.5em][l]{\rotatebox{30}{ #1}}}

\newcommand{\bsplus}{{%
  \fontsize{12}{12}\selectfont$\oplus$}}

\newcommand{\bsdot}{{%
  \fontsize{12}{12}\selectfont$\odot$}}

\newcommand{\bsminus}{{%
  \fontsize{12}{12}\selectfont$\ominus$}}

\begin{table}[t]

\mycaption{Abilities of different types of assessors to handle various types of judgments. Similar levels of ability might hint at scenarios where specific types of human assessors might be replaced by~LLMs.}
\label{table-perspective-substitution}

\centering
\setlength{\tabcolsep}{4pt}
\renewcommand{\arraystretch}{1.2}
\begin{tabular}{@{} l@{\hspace{14pt}}l@{\hspace{16pt}} llll @{}}

\toprule
                                   &                & \multicolumn{4}{c}{\bfseries Type of Judgment} \\
                                                \cmidrule{3-6}
\bfseries Type of Assessor         & \bfseries Cost & \bscellR{Preference} & \bscellR{Binary} & \bscellR{Graded} & \bscellR{Explained}\kern1.5em \\
\midrule

User                               &           free &              \bsplus &          \bsplus &          \bsplus & \bsdot  \\
Expert                             &      expensive &       \bsplus\bsplus &   \bsplus\bsplus &          \bsplus & \bsplus \\
Crowdworker                        &          cheap &               \bsdot &          \bsplus &          \bsplus & \bsdot  \\
LLM \includegraphics{chatgpt-logo} &     very cheap &              \bsplus &          \bsplus &           \bsdot & \bsplus \\

\bottomrule\\[-2ex]
\multicolumn{6}{c}{Legend:\quad \bsplus\bsplus~can judge,\quad \bsplus~depends, \quad\bsdot~unknown}
\end{tabular}
\end{table}

\paragraph{LLMs' Judgment Quality}

It is yet to be understood what the benefits and risks associated with \ac{LLM} technology are. A rather similar debate was spawned more than ten years ago with the early use of crowd workers to create relevance judgments. While before, judgments were typically made by in-house experts, the very same judgment tasks were then delegated to crowd workers, with a substantial decrease in terms of quality of the judgment, compensated by a huge increase in annotated data~\cite{HalveyVillaEtAl2015}. Quality-assurance methods were developed to obtain the highest gains~\cite{DanielKucherbaevEtAl2018}. With \acp{LLM}, history may repeat itself: a huge increase in annotated data, with a decrease in terms of quality---although the specific extent  of the deterioration is still unclear. \ac{LLM}-specific quality assurance methods will need to be developed, and, even an improvement in quality is possible.
A related idea consists in allowing \acp{LLM} to learn by observing human annotators performing the task or following an active learning paradigm~\cite{SharmaKaushik2017,YuChungEtAl2022,SeungminDonghyunEtAl2022}. The \ac{LLM} starts with mild suggestions to the assessor on how to annotate documents, then it continues to learn by considering actual decisions made by the annotator and finally improving the quality of the suggestions provided. See in this regard the scenarios ``AI Assistance'' and ``Human Verification'' in Table~\ref{table-perspective-collaboration}.

In essence, we ask the question: 
\emph{For which tasks can what type of human assessor be replaced by an \ac{LLM}?}
Table~\ref{table-perspective-substitution} provides a rough view in this regard: We distinguish four types of assessors (user, expert, crowdworker, and \ac{LLM}) over four judgment tasks: preference (which document is more relevant), binary (which of the two documents is relevant), graded (distinguish more than two levels of relevance), explained (justify the relevance decision). Table~\ref{table-perspective-substitution} is useful in showing a spectrum of substitutions, but it is unsatisfactory in clarifying the role of \ac{LLM}---we are still in the early stages of development and simply do not know (\bsdot).

\paragraph{LLMs Cost}


Related to \aiAssistance{} as well as \manualVerification{}, Table~\ref{table-perspective-substitution} shows tendencies regarding the replacement or the indispensable properties of humans in judgment tasks. The table includes a ``cost'' column that will play a role in the future, but for which only relative estimates can be provided at this time. Note that there is no clear exclusion for either party.

\subsection{Manual Verification}

\paragraph{Using Multiple LLMs as Assessors}

A difference between humans and automatic assessors concerns the number of assessors. While it is possible to hire multiple human assessors to annotate documents and, possibly, resolve disagreements between annotators~\cite{FerranteFerroEtAl2017}, this is not that trivial in the automatic assessor case.
\acp{LLM} which are trained on similar corpora are likely to produce correlated answers---but we do not know whether these are correct. A possible solution to this would include the usage of different subcorpora based on different sets of documents.
This, in turn, could lead to personalized \acp{LLM}~\cite{YoonYun2017,JaechOstendorf18,WelchGuEtAl2022}, fine-tuned on data from different types of users, which would allow to auto-annotate documents directly according to a user's subjective point of view, while also helping with increasing the pool of judgments collected. While this technology is not available yet, mostly due to computational reasons, we expect it to be available in the coming years.

\paragraph{Truthfulness \& Misinformation}

An important aspect to consider when it comes to relevance judgments is factuality. Consider the question ``do lemons cure cancer?'', for which top-ranked documents may indeed discuss healing cancer with lemons. While topically relevant, the content is unlikely to be factually correct. The result can  therefore be defined as not relevant to correctly answering the information need. To overcome this issue, human assessors have to access external information (as well as their own acquired knowledge) to determine the trustworthiness of a source as well as the truthfulness of a document.

In the fully automatic setting, we rely entirely on \acp{LLM} to verify the source and the truthfulness of the document content. This raises questions: Can we automatically assess the reliability of \ac{LLM}-generated results? Can we automate fact-checking, for example, by identifying the information source of a generative model and verifying that it is presented accurately? Going forward, it will also be vital to be able to distinguish between human-generated and \ac{LLM}-generated data, especially in contexts such as journalism where the correctness of facts is critical.

\paragraph{Bias}

\acp{LLM} are biased, the evaluation should not be.
\citet{BenderGebruEtAl2021} highlight limitations associated with \acp{LLM}, identifying a severe risk in their internal bias. \acp{LLM} are intrinsically biased \cite{BastaCostajussaEtAl2019,KuritaVyasEtAl2019,HutchinsonPrabhakaran2020} and such bias may also be reflected in the relevance judgments. For example, an \ac{LLM} might be prone to consider documents written in scientific language as relevant, while being biased against documents written in informal language.
The community should focus on finding a way to evaluate the model itself in terms of bias, and verify that, even though a model has been trained on biased data, the evaluation is not unduly affected by the same biases.

\paragraph{Faithful Reasoning.}
LLMs can generate text that contains inaccurate or false information (i.e., hallucinate). This text is often presented in such an affirmative manner that it makes it difficult for humans to detect errors. In response, the NLP community is exploring a new research direction called ``faithful reasoning''~\cite{CreswellShanahan2022}. This approach aims to generate text that is less opaque, also describing explicitly the step-by-step reasoning, or the ``chain of thoughts''~\cite{LyuHavaldarEtAl2023}.


\paragraph{Explain Relevance to LLMs}

Judgment guidelines provide a comprehensive overview of what constitutes a relevant document for a specific task---most famously, Google's search quality rating guidelines for web search have been more than 170 pages long.%
\footnote{\url{https://guidelines.raterhub.com/searchqualityevaluatorguidelines.pdf}}
It is an open question how to ``translate'' such guidelines for \acp{LLM}. 

In addition, for many tasks, relevance may go beyond topical relevance~\cite{Saracevic1996}. Sometimes, a certain style is desired. Sometimes, the truthfulness of the information is very important. Sometimes, desired information should allure users from certain communities and cultures with different belief systems. We do not yet know to what extent \acp{LLM} are capable of assessing these very different instantiations of relevance. We believe that, to properly support widely different tasks, human intervention needs to be plugged into the collection and judgment of additional facts and document aspects not yet easily discernable for an \ac{LLM}.


\subsection{Fully Automated}

\paragraph{LLM-based Evaluation of LLM-based Systems}
In the fully automated scenario, a circulatory problem can arise: How is this ranking evaluation different from being an approach that produces a ranking? In practical settings, we expect the \ac{LLM} used for ranking to be much smaller (more cost effective, lower latency, etc. achieved for example by knowledge distillation) than the \ac{LLM} used for judging. In addition, the judging \ac{LLM} can be endowed with additional information about relevant facts/questions/nuggets that the system under evaluation does not have access to. Lastly, we point to an ensemble of judging \acp{LLM} as a potential way forward.

\paragraph{Moving Beyond Cranfield}

Many assumptions and decisions taken in the relevance judgment process enable us to make the manual judgment feasible within a limited time and monetary budget. For example, we consider collections static, and relevance judgments to not change over time (a simplification as seen in \cite{Schamber1994,Voorhees2000}); we assume that the relevance of a document is not dependent on the other documents in the same ranking and that creating relevance judgments for a small set of queries provides us with a sufficiently good amount of data to compare a set of search systems with each other. If \acp{LLM} would perform reliably with little human verification, many of these assumptions could be relaxed. For example, in TREC CAsT~\cite{DaltonXiong2020,DaltonXiong2020b}\footnote{TREC CAsT is a shared task that aims at evaluating conversational agents. TREC CAsT provides information needs in the form of multi-turn conversations, each containing several utterances that a user might pose to a conversational agent.}, information needs are developing (instead of static) as the user learns more about the domain.
Hence a tree of connected information needs is defined, where one conversation takes a path through the tree. The Human-Machine evaluation paradigm might make it feasible to assess more connected (and hence, realistic) definitions of relevance.

\paragraph{Moving Beyond Human.}\label{sec:beyond-human}
Finally, we point out that there is room beyond our proposed spectrum: this point is reached when machines surpass humans in the relevance judgment task. We have witnessed this phenomenon in a variety of NLP tasks, such as scientific abstract classification~\cite{GohCaiEtAl2020}
and sentiment detection~\cite{WeismayerPezenkaEtAl2018}. Humans are likely to make mistakes when annotating documents and are limited in the time dedicated to judgment. It is likely that \acp{LLM} will be more self-consistent, and (with sufficient monetary funds) capable of providing a large number and more consistent judgments. However, if we use human-annotated data as a gold standard, we will not be able to detect when the \ac{LLM} surpasses human performance. We then will have reached the limit of measurement: We will not be able to use differences between the current evaluation paradigms to evaluate such models.

\section{Preliminary Assessment} \label{sec:evaluation}


To provide a preliminary assessment of today's \ac{LLM} capability for relevance judgments, we conducted an empirical comparison between human and \ac{LLM} assessors. This comparison includes two test collections (TREC-8 adhoc retrieval~\cite{VoorheesHarman1999} and the TREC~2021 Deep Learning Track~\cite{craswell2022overview}), two types of judgments (binary and graded), two tailored prompts and two models (GPT-3.5 and YouChat). The experiments we report in this section were conducted in January and February~2023.

\subsection{Methodology}

We want to emphasize that the experiments we present are not meant to be exhaustive, instead the goal is to explore where \acp{LLM} agree or disagree with manual relevance judgments.

\paragraph{Corpora}
We base our experiments on two test collections: (i)~the \emph{passage retrieval task} of the TREC~2021 Deep Learning Track (TREC DL~2021)~\cite{craswell2022overview}, and (ii)~the \emph{adhoc retrieval task} of TREC-8~\cite{VoorheesHarman1999}.
Besides having a large number of relevance judgments, these collections also have contrasting properties.
The TREC DL-2021 test collection comprises short documents and queries phrased as questions; the TREC-8 adhoc test collection comprises much longer, complete documents, with detailed descriptions of information needs, explicitly stating what is and is not considered relevant.
As an experimental corpus, TREC DL~2021 provides the additional benefit that its creation date falls after the time that training data was crawled for the main GPT-3.5 LLM model we are employing in our experiments (up to June 2021) but falls before the release of the model itself (November 2022)\footnote{\url{https://platform.openai.com/docs/models/overview}}. The LLM was not directly trained on TREC-DL topics and relevance judgments, nor was it used as a component in any system generating experimental runs.

\paragraph{Sampling}
Given the available relevance judgments created by professional TREC assessors, we sampled $n=1000$ TREC-8 and TREC-DL~2021 topic--document pairs from the published relevance judgments files, respectively. Due to the limited scalability of using YouChat, we restricted ourselves to $100$~samples per relevance grade for both tasks. We sampled random pairs from all available pairs, so that each relevance grade (binary for TREC-8 and graded for TREC-DL~2021) appeared with the same frequency in our sample. 

\paragraph{\acp{LLM}}

We selected two \acp{LLM} for our experiments: GPT-3.5, more specifically {\tt text-davinci-003}\footnote{\url{https://spiresdigital.com/new-gpt-3-model-text-davinci-003}}, as accessed via OpenAI's~API,%
\footnote{\url{https://platform.openai.com/docs/api-reference/introduction}} and YouChat, both in February~2023. The former is an established standard model for many applications and thus serves as a natural starting point and first baseline, the latter has been recently integrated with the You~search engine\footnote{\url{https://you.com}.} as one of the first LLMs to be fully integrated with a commercial search engine for the task of generating a new kind of search engine result page~(SERP) that resembles a Wikipedia article, where the text is a query-biased summary of the top-$k$ most relevant web pages, $k \lesssim 5$, according to You's retrieval model with numbered references to the $k$~web pages, which are listed as $k$~``blue links'' below it. The YouChat release followed closely in the wake of that of OpenAI's ChatGPT. We chose the former due to it being an order of magnitude faster and more stable at the time of writing, whereas the latter had long time spans of unreachability and instability.

\paragraph{Prompts}
We created two simple and straightforward prompts for the two corpora as shown in Figure~\ref{fig:prompt}. We explicitly did not spend time on optimizing the prompts (so-called ``prompt engineering'') to determine whether those small differences in phrasing have an impact. Rather, we kept the prompts straightforward and to the point to establish a first baseline, and leave studying the importance of the prompt for future work.

\paragraph{Answer Parsing}
We recorded each model's generated answers and translated them into binary relevance judgments. In the case of GPT-3.5, the prompts and the setting $\textit{temperature}=0$ were sufficient to constrain the model to emit only the relevance grades requested in the prompt. In the case of YouChat, with two exceptions, the answers for TREC-DL~2021 were entirely homogeneous, and started with either \emph{``The passage is relevant~[\ldots]''} or \emph{``The passage is not relevant~[\ldots]''} and were thus straightforward to parse. The answers for the TREC-8 prompts were similarly homogeneous.

\begin{figure}
\begin{greyQuote}
Instruction: You are an expert assessor making TREC relevance judgments. You will be given a TREC topic and a portion of a document. If any part of the document is relevant to the topic, answer ``Yes''. If not, answer ``No''. Remember that the TREC relevance condition states that a document is relevant to a topic if it contains information that is helpful in satisfying the user's information need described by the topic. A document is judged relevant if it contains information that is on-topic and of potential value to the user.
\newline
\newline
Topic: \texttt{\{topic\}}\newline
Document: \texttt{\{document\}}\newline
Relevant?
\end{greyQuote}
\vspace{0.25cm}
\begin{greyQuote}
Instruction: Indicate if the passage is relevant for the question.
\newline
\newline
Question: \texttt{\{question\}}\newline
Passage: \texttt{\{passage\}}
\end{greyQuote}
\caption{Prompts used in our \S\ref{sec:evaluation} experiments on TREC-8 (top) and TREC-DL~2021 (bottom). The placeholders \texttt{\{topic\}} and \texttt{\{document\}} (TREC-8) and \texttt{\{question\}} and \texttt{\{passage\}} (TREC-DL~2021) are replaced with our sampled pairs.}
\label{fig:prompt}
\end{figure}

\subsection{Results}

In Table~\ref{tbl:trec8} we report our results for TREC-8 assessors vs.\ GPT-3.5 and YouChat respectively. We observe a clear divide according to the relevance label: for the documents judged by human assessors as non-relevant, GPT-3.5 generates the same answer in~90\% of the cases. In contrast though, for the documents judged as relevant by human assessors, this agreement drops to~50\%. Likewise, YouChat has judged 74\% of the non-relevant correctly to be non-relevant, whereas this agreement drops even more to 33\% for the relevant documents.

\begin{table}[tb]
\mycaption{Judgment agreement on TREC-8 between TREC assessors and the~LLMs; 1000~topic--document pairs for GPT-3.5 and 100~for each grade (relevant, non-relevant) for YouChat.}
\label{tbl:trec8}
\begingroup
\begin{tabular}{@{}llcc@{}c}
\toprule
\textbf{LLM} & \textbf{Prediction} & \multicolumn{2}{c@{}}{\textbf{TREC-8 Assessors}} & \textbf{Cohen's} $\boldsymbol{\kappa}$\\
\cmidrule(l@{\tabcolsep}){3-4}
&              & Relevant & Non-relevant & \\
\midrule
\multirow{2}{*}{GPT-3.5}
& Relevant     & \textbf{237}      & \phantom{0}48 & \multirow{2}{*}{0.38}\\
& Non-relevant & 263      &  \textbf{452} \\
\midrule
\multirow{2}{*}{YouChat}
& Relevant     & \phantom{0}\textbf{33}     & \phantom{0}26 & \multirow{0}{*}{0.07} \\
& Non-relevant & \phantom{0}67     & \phantom{0}\textbf{74} \\
\bottomrule
\end{tabular}
\endgroup
\end{table}

Interestingly though, when we consider the results of our second experiment in Table~\ref{tbl:dl21}---TREC-DL21 assessors vs.\ GPT-3.5 and YouChat respectively---the picture changes completely. We observe almost the opposite of what we have just described in the previous paragraph. Concretely: the higher the relevance grade, the more YouChat is in line with the human assessors. For~96 out of~100 question--passage pairs that TREC assessors judges as highly relevant (i.e.,~relevance grade~3), YouChat agreed with the assessor. In contrast, for the non-relevant question--passage pairs, the agreement is random. YouChat only agrees with manual assessments on~42 of~100 non-relevant question--passage pairs.

\begin{table}[tb]
\mycaption{Judgment agreement on TREC-DL~2021 between TREC assessors and the~LLMs; 100~question--passage pairs for each grade from 3~(highly relevant) to 0~(non-relevant).}
\label{tbl:dl21}
\begingroup
\begin{tabular}{@{}llcccc@{}c}
\toprule
\textbf{LLM} & \textbf{Prediction} & \multicolumn{4}{c@{}}{\textbf{TREC-DL~2021 Assessors}}  & \\
\cmidrule(l@{\tabcolsep}){3-6}
&                     &  3 &  2 &  1 &  0 &  \textbf{Cohen's} $\boldsymbol{\kappa}$\\
\midrule
\multirow{2}{*}{GPT-3.5}
& Relevant            & \textbf{89} & \textbf{65} & \textbf{48} & 16 & \multirow{2}{*}{0.40}\\
& Non-relevant        & 11 & 35 & 52 & \textbf{84}  \\
\midrule 
\multirow{2}{*}{YouChat}
& Relevant            &  \textbf{96} &  \textbf{93} & \textbf{79} & 42 & \multirow{2}{*}{0.49} \\
& Non-relevant        & \phantom{0}4 & \phantom{0}7 & 21 & \textbf{58} \\
\bottomrule
\end{tabular}
\endgroup
\end{table}

As a possible explanation for these observations, we hypothesize that human assessors are better at recognizing subtle details that distinguish relevant from non-relevant documents.
 When exploring coarse-grained graded relevance judgments, however, LLMs demonstrate a better correlation with the human judgments. We suspect that LLMs would be helped by symmetrically centering relevance judgments around 0~(``borderline relevant'') with a range from~-3 to~3.

\section{Re-Judging TREC 2021 Deep Learning} \label{sec:charlie}

To complement the experiments reported in Section~\ref{sec:evaluation}, in this section we report an experiment to fully re-judge submissions to a single evaluation exercise, the passage ranking task of the TREC~2021 Deep Learning Track~\cite{craswell2022overview}. Unlike the experiments reported in Section~\ref{sec:evaluation}, which focused on binary relevance, we attempt to adhere as closely as possible to the methodology used in the track itself, including the use of graded judgments.

\subsection{Methodology}

\citet{craswell2022overview} provide full details of the passage ranking task of the TREC~2021 Deep Learning Track (TREC-DL~2021). TREC-DL~2021 track participants submitted a total of 63~experimental runs, with each run comprising up to 1000~ranked passages for 200~test queries. These runs were pooled, and 53~queries were judged by assessors using a combination of methods, including active learning~\cite{trec2021,AbualsaudGhelaniEtAl2018}. This generated a total of 10,828~judgments on a 4-point scale: ``Perfectly relevant'' $\succ$ ``Highly relevant'' $\succ$ ``Related'' $\succ$ ``Irrelevant''.

We re-judged this pool using the GPT-3.5 {\tt text-davinci-003} language model, as accessed through Open AI's~API in February~2023. Consistent with a classification task---and consistent with the GPT-3.5 experiments reported in Section~\ref{sec:evaluation}---we set the \emph{temperature} parameter to~0, but otherwise default parameters and settings. 

Since our prompt is relatively long, we provide it online.%
\footnote{\url{https://plg.uwaterloo.ca/~claclark/trec2021_DL_prompt.txt}}
The prompt is inspired by a prompt appearing in \citet{Ferraretto23}: importantly---and different from the prompt in Figure~\ref{fig:prompt}---it leverages few-shot learning by listing  multiple \emph{examples} illustrating different levels of relevance for different queries. We provide one example each for ``Perfectly relevant'', ``Highly relevant'', and ``Related''; we provide two examples for ``Irrelevant'', with one providing a judged ``Irrelevant'' passage, and the other providing an unrelated passage from the pool. These examples were chosen arbitrarily from the pool, based on the TREC judgments. We also used the term ``Relevant'' in the prompt, instead of ``Related'', since ``Related'' is a non-standard label for relevance judgments; in preliminary experiments, the LLM would sometimes return ``Relevant'' unprompted. Using this prompt, judgments cost around USD~1~cent each. For this experiment we spent a total of USD~111.90, including a small number of duplicate requests due to failures and other issues. To provide a basis for comparison, \citet{ClarkeVtyurinaEtAl2021} report spending USD~25 cents per human label on a judgment task of similar scope---with a single-page ``prompt'' and no training of assessors.

\subsection{Results}

\begin{figure}[t]%
\centering%
\includegraphics[width=0.99\linewidth, keepaspectratio]{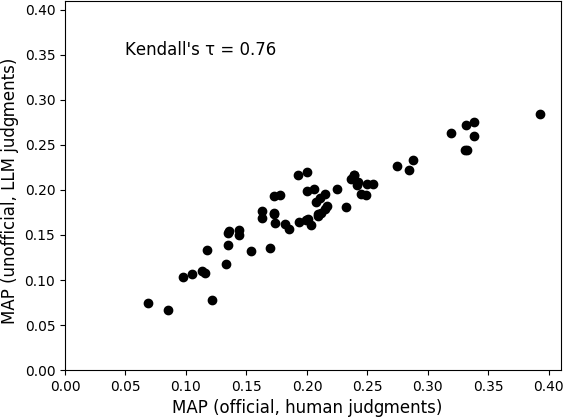}
\includegraphics[width=0.99\linewidth, keepaspectratio]{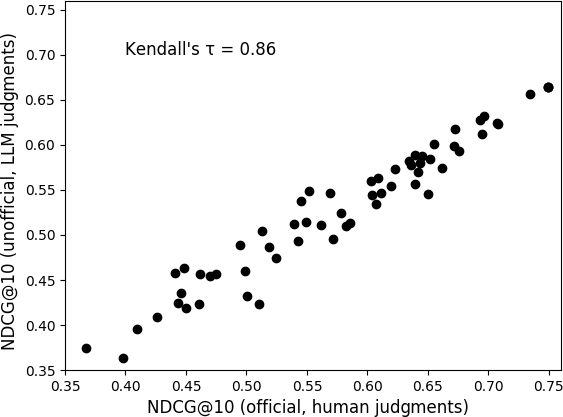}%
\caption{Scatter plots comparing the performance of TREC 2021 Deep Learning Track passage ranking runs using official, human judgments and unofficial, LLM judgments, with MAP (top) and NDCG@10 (bottom). A point represents the performance of a single experimental run avg. over all queries.}%
\label{fig:DL}%
\end{figure}

\begin{table}
\mycaption{Confusion matrices comparing all official TREC question--passage judgments with GPT-3.5 judgments on TREC-DL~2021 question--passage pairs. The upper matrix (GRADED) compares judgments on all four relevance levels. The lower matrix (BIN.) collapses the relevance labels to two levels, following the TREC-DL~2021 convention for computing binary measures.} 
\label{tbl:DL}

\begin{tabular}[t]{@{}llcccr@{}}
\toprule
& \textbf{Prediction} & \multicolumn{4}{c}{\textbf{TREC-DL~2021 Assessors}}\\
\cmidrule(l@{\tabcolsep}){3-6}
&& Perf.\ rel. & High.\ rel. & Related & Irrel. \\
\midrule
\multirow{4}{*}{\begin{sideways}\textbf{GRADED}\end{sideways}}
& Perfectly relevant & \textbf{250} & 248 &\ 177 &\ \ 87 \\
& Highly relevant    & 360 & \textbf{575} &\ 628 &\ 370 \\
& Relevant           & 328 & 880 &\ \textbf{798} &\ 442 \\
& Non-relevant         & 148 & 638 & 1460 & \textbf{3439} \\
\bottomrule
\\
\bottomrule
& \textbf{Prediction} & \multicolumn{4}{c}{\textbf{TREC-DL~2021 Assessors}}\\
\cmidrule(l@{\tabcolsep}){3-6}
&& \multicolumn{2}{c}{Relevant} & \multicolumn{2}{c}{Not relevant} \\
\midrule
\multirow{2}{*}{\begin{sideways}\textbf{BIN.}\end{sideways}}
& Relevant     & \multicolumn{2}{c}{\textbf{1433}} & \multicolumn{2}{c}{1262} \\
& Non-relevant & \multicolumn{2}{c}{1994} & \multicolumn{2}{c}{\textbf{6139}} \\
\bottomrule
\end{tabular}
\end{table}

Table~\ref{tbl:DL} provides a summary of the results. We provide a summary for both the full 4-point relevance scale and a binary relevance scale, which follows the TREC-DL~2021 convention for computing binary measures such as MAP. This convention maps ``Perfectly relevant'' and ``Highly relevant'' to ``Relevant'', and maps ``Relevant'' and ``Irrelevant'' to ``Not relevant''. Whereas for Table~\ref{tbl:dl21}, in order to compare results with YouChat, we followed the more usual convention of treating all grades except ``Irrelevant'' as relevant. On the binary judgments of Table~\ref{tbl:DL}, Cohen's~$\kappa = 0.26$,  a level of agreement that is conventionally described as ``fair''. To provide a basis for comparison, \citet{cormack1998efficient} report results corresponding to a Cohen's~$\kappa = 0.52$ on a similar experiment comparing two types of human judgments, a level of agreement conventionally described as ``moderate''.

We applied the LLM~judgments to compute standard evaluation measures on the runs submitted to TREC-DL~2021, with the results shown in Figure~\ref{fig:DL}. Kendall's~$\tau$ values show the correlation between system rankings. To provide a basis for comparison, \citet{Voorhees2000} report a Kendall's~$\tau = .90$ for MAP on a similar experiment comparing two types of human judgments. Nonetheless, the top run under the official judgments remains the top run under the LLM~judgments.

We find that measures computed under the LLM~judgments are less sensitive than measures computed under human judgments.
Sensitivity (or ``discriminative power'') measures the ability of an evaluation method to recognize a significant difference between retrieval approaches~\cite{sakai06,ymt18,ClarkeVtyurinaEtAl2021}.
To compute sensitivity, we take all pairs of experimental runs and compute a paired t-test between them.
A pair with~$p < 0.05$ is considered to be \emph{distinguished}~\cite{ymt18}, with sensitivity defined as $\frac{\textit{\# of distinguished pairs}}{\textit{total pairs}}$.
Since we do not correct for the multiple comparisons problem, some of the distinguished pairs may not represent actual significant differences. 
%
Under human judgments 72\% of systems are distinguished under MAP (74\% under NDCG@10). In contrast, under GPT-3.5 judgments only 65\% are distinguished (69\% under NDCG@10). 



\section{Perspectives for the Future} \label{sec:future}

As this is a perspectives paper, we now provide two opposing perspectives---for and against the use of \acp{LLM} for automatic relevance judgments---and a compromise perspective, all of which are informed by our analysis of the literature, our experimental evidence, and our experience as \ac{IR} researchers.

\subsection{In Favor of Using LLMs for Judgments}

More than just the plain judgment of relevance, LLMs are able to produce a natural language explanation \textit{why} a certain document is relevant or not to a topic~\cite{Ferraretto23}. Such AI-generated explanations may be used to assist human assessors in relevance judgments, particularly non-experts like crowdworkers. This setup may lead to better quality judgments as compared to the unsupported crowd. While LLM-generated labels and explanations may bias human assessors and mislead them on the relevance a document has, human assessors may serve as a quality control mechanism for the LLM as well as a feedback loop for the LLM to continuously improve its judgments. 
Our pilot experiments demonstrate that it is feasible for LLMs to indicate when a document is likely not relevant. We might therefore let human annotators assess (a)~first those documents that are deemed relevant by~LLMs, or (b)~a subsample of documents from those considered relevant by the LLM, as an LLM can be run at scale. Thereby, we envision the use of LLMs to reduce annotation cost/time when creating high-quality IR~evaluation collections.

Noteworthy, LLMs have actual conceptual advantages over humans when it comes to a fair and consistent judgment. They can judge the relevance of documents without being affected by documents they have seen before, and with no boredom or tiredness effects. They can also ensure to treat conceptually identical documents identically. At the same time, they will often have seen much more information on a specific topic than a human.
Another advantage of today's LLMs is their inherent ability to process and generate text in many different languages. For multilingual corpora (which often appear in industrial settings) the assessment is typically restricted to a small subset of languages due to the limited availability of assessors. With LLMs as assessment tool, this limitation no longer applies.

LLMs are not just restricted to one input modality and thus conducting assessments that require the simultaneous consideration of multiple pieces of content (e.g. judging a web page based on the text but also the document's structure, visual cues, embedded video material, etc.) at the same time becomes possible. 
Finally, we note the cost factor---if we are able to judge hundreds of thousands of documents for a relatively small price, we can build much larger and much more complex test collections with regularly updated relevance assessments, in particular in domains that today lack meaningful test collections.


In summary, LLMs can provide explanations, scalability, consistency, and a certain level of quality when performing relevance judgments, underlining the great potential of deploying them as a complement to human assessors in certain judgments task.

\subsection{Against Using LLMs for Judgments}

While we have given several reasons to believe that we are close to using \acp{LLM} for automatic relevance judgment, there are also several concerns that should be addressed by the research community before being able to deploy full-fledged automatic judgment.
The primary concern is that \acp{LLM} are not people. \ac{IR} measures of effectiveness are ultimately grounded in a human user's relevance judgment.
Relevance is subjective, and changes over time for the same person \cite{mizzaro1997relevance}.
Even if \acp{LLM} are increasingly good at mimicking human language in evaluating contents, jumping from that up to trusting the model as if it were a human is a big leap of faith. Currently, there is no proof that the evaluation made by \acp{LLM} has any relationship to reality.
This raises an essential question: \emph{If the output from an \ac{LLM} is indistinguishable from a human-made relevance judgment, is this just a distinction without a difference?} After all, people disagree on relevance and change their opinions over time due to implicit and explicit learning effects. Usually, however, those disagreements do not have an effect on the evaluation unless there are systematic causes~\cite{Voorhees2000,BaileyCSTVY08}.  
To safely adopt \acp{LLM} to replace human annotators, the community should examine whether \ac{LLM}-based relevance judgments may in fact be systematically different from those of real users. Not only do we know this affects the evaluation, but the complexity (or black-box nature) of the model precludes defining systematic bias in any useful way. 

\medskip

There is a general concern about solely evaluating IR research with relevance assessment: Information retrieval systems are not just result-ranking machines, but are a system that is to assist a human to obtain information. Hence, only the user who consumes the results could tell which ones are useful.


\medskip

Another concern of applying \acp{LLM} as relevance annotators regards the ``circularity'' of the evaluation. Assume we are able to devise an annotation model based on \acp{LLM}. The same model could ideally also be used to retrieve and rank documents based on their expected relevance. If the model is used to judge relevance both for annotation and for retrieval, its evaluation would be overinflated, possibly with perfect performance. Vice-versa, models based on widely different rationales (such as BM25 or classical lexical approaches), might be penalized, because of how they estimate document relevance. 
As counter-considerations, we might hypothesize that the model used to label documents for relevance (a)~is highly computationally expensive, making it almost unfeasible to use it as a retrieval system, and/or (b)~has access to more information and facts than the retrieval model. The former holds as long as we do not use the automatic annotator as an expensive re-ranker capable of dealing with just a few documents. The latter, on the other hand, does not solve the problem of the automatic annotation, but simply moves it: Either, the additional facts and information need to be annotated manually; then the human annotator remains essential. Or, the facts can be collected automatically; then we may assume that also a retrieval system could obtain them.

Other concerns arise if we even consider generative models as a replacement for traditional~IR and search. In a plain old search engine, results for a query are ranked according to predicted relevance (ignoring sponsored results and advertising here). Each has a clear source, and each can be inspected directly as an entity separate from the search engine. Moreover, users frequently reformulate queries and try  suggestions from the search engine, in a virtuous cycle wherein the users fulfill or adjust their conceptual information needs. Currently, hardly any of these is possible using LLM-generated responses: The results often are not attributed, rarely can be explored or probed, and are often wholly generated. Also, ``prompt engineering'' is still explored much less and hence more opaque than query reformulation. LLMs~will not be usable for many information needs until they can attribute sources reliably and can be interrogated systematically. We expect working solutions to these issues to be just a matter of time, though.  

Finally, there are significant socio-technical concerns. Generative AI models can be used to generate fake photos and videos, for extortion purposes, or for misinformation. They are perceived as stealing the work of others. 
Furthermore, \acp{LLM} are affected by bias, stereotypical associations~\cite{BastaCostajussaEtAl2019,KuritaVyasEtAl2019}, and adverse sentiments towards specific groups~\cite{HutchinsonPrabhakaran2020}. 
Critically, we cannot assess whether the LLM may have seen information that biases the relevance judgment in an unwanted way, let alone that the company owning the LLM may change it anytime without our knowledge or control. 
As a result, we ourselves as the authors of this perspectives paper disagree on whether, as a profession and considering the ACM's Code of Ethics, we should use generative models in deployed systems {\it at all} until these issues are worked out.
%
%
%


\subsection{A Compromise: Double-checking LLMs and Human--Machine Collaboration}


Our pilot study in Sections~\ref{sec:evaluation} and \ref{sec:charlie} finds a reasonable correlation between highly-trained human assessors and a fully automated LLM, yielding similar leaderboards. This suggests that the technology is promising and deserves further study. The experiment could be implemented to double-check LLM judgments: 
Produce fully automated as well as human judgments on a shared judgment pool, then analyze correlations of labels and system rankings,
then decide whether LLM's relevance judgments are good enough to be shared as an alternative test collection with the community. The automatic judgment paradigm should be revealed along with prompts, hyperparameters, and details for reproducibility.
We also suggest to declare which judgment paradigm was chosen when releasing data resources (such as in TREC CAR). At the very least, such automatic judgments could be used to evaluate early prototypes of approaches, for initial judgments for novel tasks, and for large-scale training.

\medskip

While the discussion is easily dominated by fully automated evaluation---these are merely an extreme point on our spectrum in Section \ref{sec:spectrum}. The majority of authors do not believe this constitutes the best path towards credible IR research. For example, ``AI Assistance'' is probably the most credible path for \acp{LLM} to be incorporated during evaluation. However, it is also the least explored so far.

This calls for more research on innovative ways to use~\acp{LLM} for assistance during the judgment process and how to leverage humans for verifying the \acp{LLM}'~suggestions. As a community, we should explore how the performance of human assessors changes, when they are shown rationales or chain-of-thoughts that are generated by \acp{LLM}. Human assessors often struggle to see a pertinent connection when they are lacking world knowledge. An example of this issue is the task of assessing the relevance of ``diabetes'' for the topic ``child trafficking''. \acp{LLM} can generate rationales that can explain such connections. However, it requires a human to realize when such a rationale was hallucinated. Only a human can assess whether the information provided appears true and reliable.

\balance

\section{Conclusion} \label{sec:conclusion}


In this paper, we investigated the opportunity that large language models (\acp{LLM}) now provide to generate relevance judgments automatically. We discussed previous attempts to automatize and scale-up the relevance judgment task, and we presented experimental results showing promise in the ability to mimic human relevance assessments. Finally, we presented our views on why and why not the research community should employ \acp{LLM} in some fashion in the IR evaluation process. Undoubtedly, more research on \acp{LLM} for relevance judgment is to be carried out in the future, for which this paper provides a starting point.
\begin{acks}
This paper is based on discussions during a breakout group at the Dagstuhl Seminar 23031 on ``Frontiers of Information Access Experimentation for Research and Education''. We express our gratitude to the Seminar organizers, Christine Bauer, Ben Carterette, Nicola Ferro, and Norbert Fuhr.

\smallskip


Certain companies and software are identified in this paper in order to specify the experimental procedure adequately. Such identification is not intended to imply recommendation or endorsement of any product or service, nor is it intended to imply that the software or companies identified are necessarily the best available for the purpose.

\smallskip
This material is based upon work supported by the National Science Foundation under
Grant No. 1846017. Any opinions, findings, and conclusions or recommendations expressed in this material
are those of the author(s) and do not necessarily reflect the views of the National Science
Foundation.

\end{acks}

\balance
\bibliographystyle{ACM-Reference-Format}
\bibliography{biblio,dietz}



\end{document}